\documentclass[fleqn,usenatbib]{mnras}

\usepackage{newtxtext,newtxmath}

\usepackage[T1]{fontenc}

\DeclareRobustCommand{\VAN}[3]{#2}
\let\VANthebibliography\thebibliography
\def\thebibliography{\DeclareRobustCommand{\VAN}[3]{##3}\VANthebibliography}

\usepackage{graphicx}	
\usepackage{amsmath}	
\usepackage{amssymb}	

\title[X-ray photolysis of CH$_3$COCH$_3$ ice]{X-ray photolysis of CH$_3$COCH$_3$ ice: Implications for the radiation effects of compact objects towards astrophysical ices}

\author[G. A. Carvalho and S. Pilling]{
G. A. Carvalho,$^{1}$\thanks{E-mail: geanderson.araujo.carvalho@gmail.com}
and S. Pilling,$^{1}$
\\
$^{1}$Instituto de Pesquisa e Desenvolvimento (IP\&D), Universidade do Vale do Para\'iba (UNIVAP), Av. Shishima Hifumi 2911,\\ S\~ao Jos\'e dos Campos, SP, CEP 12244-000, Brazil
}

\date{Accepted XXX. Received YYY; in original form ZZZ}

\pubyear{2015}

\begin{document}
\label{firstpage}
\pagerange{\pageref{firstpage}--\pageref{lastpage}}
\maketitle

\begin{abstract}
 In this study, we employed broadband X-rays ($6-2000$ eV) to irradiate the frozen acetone CH$_3$COCH$_3$, at the temperature of 12 K, with different photon fluences up to $2.7\times 10^{18}$ photons cm$^{-2}$. Here, we consider acetone as a representative complex organic molecule (COM) present on interstellar ice grains. The experiments were conduced at the Brazilian synchrotron facility (LNLS/CNPEN) employing infrared spectroscopy (FTIR) to monitor chemical changes induced by radiation in the ice sample. We determined the effective destruction cross-section of the acetone molecule and the effective formation cross-section for daughter species. Chemical equilibrium, obtained for fluence $2\times 10^{18}$ photons cm$^{-2}$, and molecular abundances at this stage were determined, which also includes the estimates for the abundance of unknown molecules, produced but not detected, in the ice. Timescales for ices, at hypothetical snow line distances, to reach chemical equilibrium around several compact and main-sequence X-ray sources are given. We estimate timescales of 18 days, 3.6 and 1.8 months, $1.4\times 10^9-6\times 10^{11}$ years, 600 and $1.2\times 10^7$ years, and $10^7$ years, for the Sun at 5 AU, for O/B stars at 5 AU, for white dwarfs at 1 LY, for the Crab pulsar at 2.25 LY, for Vela pulsar at 2.25 LY, and for Sagittarius A* at 3 LY, respectively. This study improves our current understanding about radiation effects on the chemistry of frozen material, in particular, focusing for the first time, the effects of X-rays produced by compact objects in their eventual surrounding ices.
\end{abstract}

\begin{keywords}
 Sun: X-rays -- White dwarfs -- Neutron stars -- Black holes -- Stars: X-rays -- Interstellar medium
\end{keywords}



\section{Introduction}

Acetone is an interstellar and cometary molecule, first discovered in the interstellar medium (ISM) by \cite{Combes1987Jun} inside the Sgr B2 molecular cloud and posteriorly confirmed by \cite{Snyder2002Oct}. Acetone in outer space was also observed in the star-forming region Orion-KL and comets \citep{Friedel2005oct,Goesmann2015jul}. The aforementioned observations of acetone in gas phase in comets and ISM indicate that acetone may be found in space also in its ice phase, since gas phase is often linked with the ice one through sublimation in cold environments and inside dense molecular clouds.

\cite{Walsh2014Mar} have presented a model for the abundances of complex organic molecules (COMs) in protoplanetary disks also including acetone in both gas and ice phase. From the models, the authors suggest that the abundances of acetone in the ice grains could be as large as $2\times 10^{17}$ cm$^{-2}$, at distances around 50 AU from the T Tauri-like star. \cite{Hudson2018Feb} has also suggested that studies on acetone in solid phase are important due to detection of oxide propylene in ISM, since the former could be formed from radiation-induced isomerization of the later. 

Ionizing radiation comes mainly from stellar sources and interacts with astrophysical ices enriching chemical complexity. The energy delivered by radiation induces chemical changes in the ices destroying molecules and also producing new species, such as acetone from oxide propylene. Several laboratory investigations have addressed the effects of radiation on pure acetone ice samples; for example, \cite{Hudson2018Feb} used 1 MeV protons to irradiate acetone ice samples at 20K and infrared spectroscopy to study radiation products, thus corroborating and correcting the previous work of \cite{Andrade2014Sep}. In \cite{Almeida2014Mar}, authors irradiated acetone ice samples with soft X-rays (0.01 – 2000 eV) to determine desorption yields and photodissociation cross-section.

In this manuscript we focus on the effect of X-rays produced by stellar sources (from main-sequence to compact) on its eventual circumstellar organic rich ices. We considered a sample of OB stars, white dwarfs (WD), pulsars, black holes (BH), and the X-ray flux of the Sun, also together with model dependent data for young stellar objects. Depending on the intensity of ionizing field of such X-rays sources the snow line can be closer or far away the central object, but the existence of ices in all these environments are expected in the literature \citep{Debes2019Feb,Koester2014Jun,Manser2019Apr,Yan2013May}.

Some protoplanetary nebulae presented ice-band spectra \citep{Debes2019Feb,Koester2014Jun,Manser2019Apr,Yan2013May}. Although not being full WD it indicates that ice material can be closer to those compact objects. Other indications of icy and volatile material close to white dwarfs are: observations on atmospheric pollution by heavy elements in about a quarter to a half of all white dwarfs \citep{Zuckerman2003Oct,Zuckerman2010Sep,Koester2014Jun} and the direct evidence of planetary systems orbiting around white dwarfs, such as the transit features in the light curve of WD 1145+017 and emission line profiles from the debris disk around the white dwarf SDSS J122859.93+104032.9 \citep{Manser2019Apr}. Pulsars have also shown to be orbited by planet-size bodies \citep{Bailes1991Jul,Wolszczan1992Jan} and, surprisingly, supermassive massive black holes were shown also to be capable of hosting planetary systems as indicated by \cite{Wada2019Nov}. Acetone transitions have been detected also from disk-like structures of protostars as reported by \cite{Isokoski2013Jun}.

Section \ref{meth} describes the employed methodology. Section \ref{reactions} provides a review on reaction routes used to guide our spectral assignments. Results are listed and discussed in section \ref{resul}, with emphasis on the effective destruction and production cross section, as well as at the estimate for the unknown material produced in the ice during photolysis and the equilibrium chemistry scenario. Section \ref{sectIV} presents the astrophysical implications focus on the timescale to reach chemical equilibrium in selected space environment (mainly compact objects). Conclusion and final remarks are given in section \ref{sectV}.


\section{Methodology}\label{meth}

In this work we use data obtained by the astrochemistry and astrobiology group of the Universidade do Vale do Paraíba (UNIVAP). The experimental data was acquired at the facilities of the Brazilian Synchrotron Light Laboratory (LNLS) at Campinas, Brazil. To simulate the astrophysical environments of interest we employed a high-vacuum chamber coupled to the spherical grating monochromator (SGM) beamline, which produced an X-ray broadband spectrum (from 6eV up to 2keV). Our main objective is to simulate acetone in the ice phase in environments where they could be exposed to ionizing soft X-rays, such as the surrounds of compact objects. Beamline details can be found elsewhere \citep[e.g.][]{Rodrigues1998May,Pilling2015Sep}.

Briefly, the gaseous acetone (SIGMA ALDRICH; purity 99.99\%) was deposited onto a polished, clean ZnSe substrate inside experimental chamber, previously cooled to 12 K. The deposition was carried out with 20 mbar of acetone during approximately 400 seconds. To perform {\it in-situ} chemical analyses of the samples we employ a Fourier transform infrared (FTIR) spectrometer (Agilent Inc., model Cary 630) coupled to experimental chamber. The spectra were taken in the range $4000-650$ cm$^{-1}$, with 2 cm$^{-1}$ resolution. After gas deposition an infrared spectrum was taken to characterize the virgin ice sample (e.g. location and intensities of bands, ice thickness, eventual impurities). Infrared spectrum was also collected after different radiation fluences to characterize the sample changes due to the photolysis by X-rays. The total irradiation time was around 450 min (final fluence of $2.7\times 10^{18}$ photons cm$^{-2}$). The pressure inside the vacuum chamber during irradiation processes was below $3\times 10^{-8}$ mbar. More details about the chamber and the employed methodology can be consulted in \citep[e.g.][]{Pilling2015Sep,Rachid2017Dec,deA.Vasconcelos2017Nov}.

The sample thickness was determined to be around 1.7 $\mu$m. To determine sample thickness we employ the methodology presented in \cite{Pilling2011Aug} by using the CC vibration mode (1228 cm$^{-1}$) with band strength $A=7.35\times 10^{-18}$ cm/molecule and mass density $\rho=0.763$ g/cm$^3$. The maximum photon dose is calculated dividing the maximum fluence by the ice column density at the end of experiment, and taking into account an average photon energy of 1 keV, which gives a maximum dose of $1.34\times 10^3$ ev/molecule, this value is obtained by taking acetone column density $2.02\times 10^{18}$ cm$^{-2}$ along with maximum fluence, $2.7\times 10^{18}$ photons/cm$^2$. A control on impurities and characterization of radiation beam followed the same protocol of our previous works \citep[see][]{Pilling2015Sep}. As indicated by \cite{Pilling2015Sep}, the integrated photon flux in the sample, correspondent to $6-2000$ eV, was estimated as $1.4\times 10^{14}$ photons cm$^{-2}$ s$^{-1}$, and the integrated energy flux was approximately $7.6\times 10^{14}$ erg cm$^{-2}$ s$^{-1}$. In this work, we focus the astrophysical implications on the effects of the soft X-ray in the ices in the vicinity of pre-main-sequence, main-sequence and compact objects, in places such as the ices close to protostars, and the eventual ice torus surrounding white dwarfs, neutron stars or black holes.

\section{Acetone decomposition and reaction routes}\label{reactions}

Past works on acetone decomposition in the gaseous phase have shown that decomposition is unimolecular and follows the possible schemes \citep{Rice1929Sep},
\begin{eqnarray}
  \hspace*{-2.5cm}&& {\rm CH}_3 {\rm COCH}_3 \rightarrow {\rm CO} + 2{\rm CH}_3, \\
  \hspace*{-2.5cm}&& {\rm CH}_3 {\rm COCH}_3 \rightarrow {\rm H}_2 {\rm CCO} + {\rm CH}_4. \label{ch4}
\end{eqnarray}

\cite{Smith1944Aug} proposed a different reaction chain as follows
\begin{eqnarray}
  \hspace*{-1cm}&& {\rm CH}_3 {\rm COCH}_3 \rightarrow {\rm CO} + 2{\rm CH}_3, \\
  \hspace*{-1cm}&& {\rm CH}_3 + {\rm CH}_3 {\rm COCH}_3 \rightarrow {\rm CH}_4 + {\rm CH}_2 {\rm COCH}_3,\\
  \hspace*{-1cm}&& {\rm CH}_2 {\rm COCH}_3 \rightarrow {\rm CH}_3 + {\rm H}_2 {\rm CCO},\label{ch3}\\
  \hspace*{-1cm}&& {\rm CH}_3 + {\rm CH}_2 {\rm COCH}_3 \rightarrow {\rm C}_2{\rm H}_5 {\rm COCH}_3.
\end{eqnarray}

In both cases metil radical, carbon monoxide, ketene and methane are listed as reaction products. More recently, \cite{Hudson2018Feb} have studied radiation chemistry induced on acetone ices by 1 MeV protons. \cite{Hudson2018Feb} listed detailed reaction routes and in addition to above reactions, he listed the formation of CO$_2$ and other molecules as product of reactions below
\begin{eqnarray}
  \hspace*{-1.2cm}&&{\rm H}_2 {\rm CCO} + {\rm H}_2 {\rm CCO} \rightarrow {\rm CH}_2 {\rm CCH}_2 + {\rm CO}_2, \\ 
  \hspace*{-1.2cm}&& {\rm CH}_3 {\rm COHCH}_3 + {\rm CH}_3 {\rm COCH}_3 \rightarrow {\rm CH}_3 {\rm CHOHCH}_3 + \nonumber \\ &&  {\rm CH}_2 {\rm COCH}_3.
\end{eqnarray}

Within radiation products identified by \cite{Hudson2018Feb} there were carbon monoxide and dioxide, methane, ketene and isopropanol. Molecules indicated as not favoured as radiation products were hydrocarbons, such as C$_2$H$_2$, C$_2$H$_4$ and C$_2$H$_6$. Reaction routes presented here will form the base of our assignments in next sections.

\section{Results}\label{resul}

We started our analysis by investigating the bands presented in the infrared spectra of the samples and its evolution with fluence. From that, we derived several physicochemical parameters and the molecular composition in the next sessions. In analyzing spectra we use OMNIC software to make baseline corrections and calculate peak areas.

\begin{figure*}
\centering
\includegraphics[height=7cm, width=0.80\textwidth]{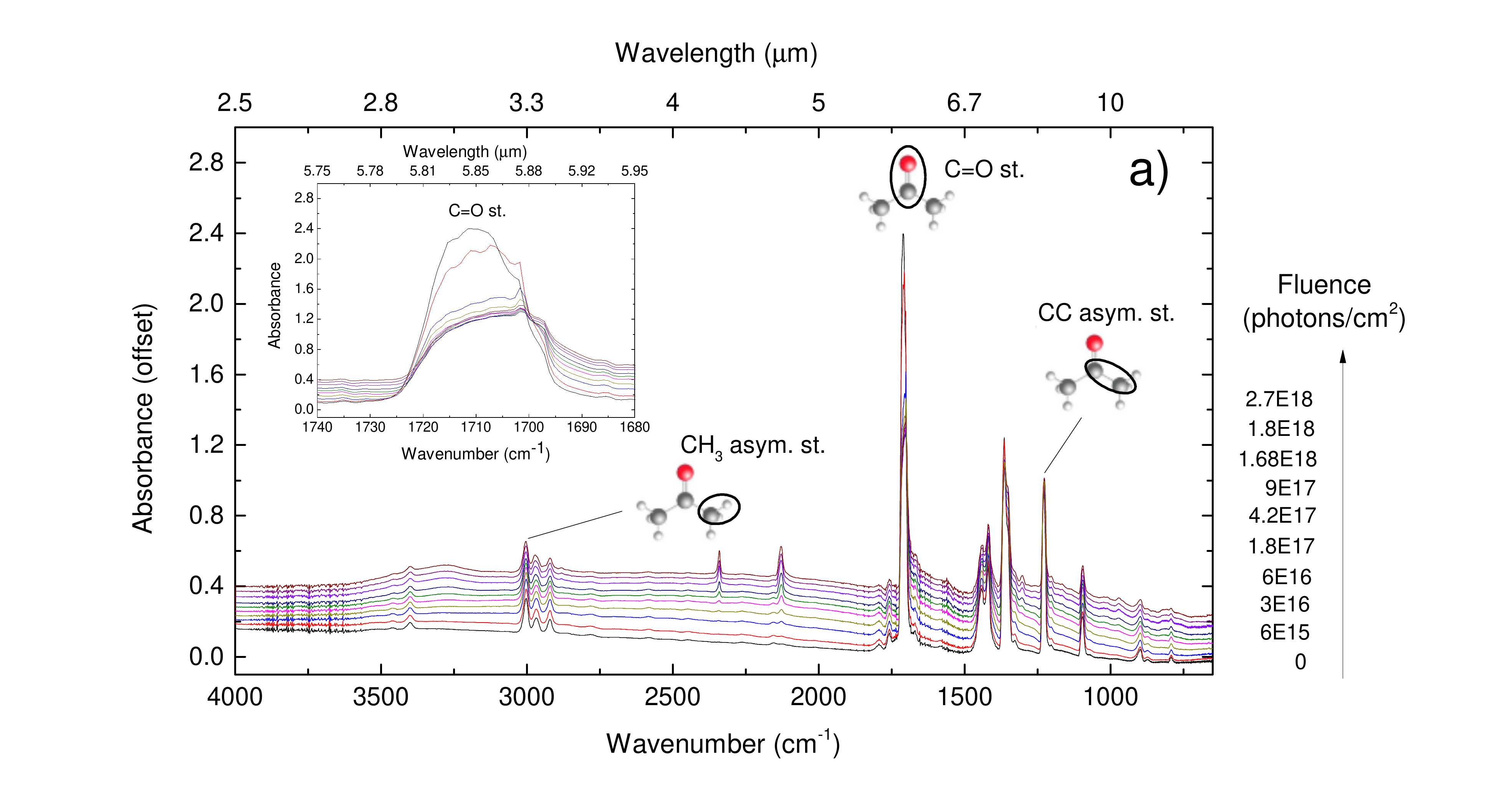}
\vspace*{-0.1cm}
\includegraphics[height=7cm, width=0.80\textwidth]{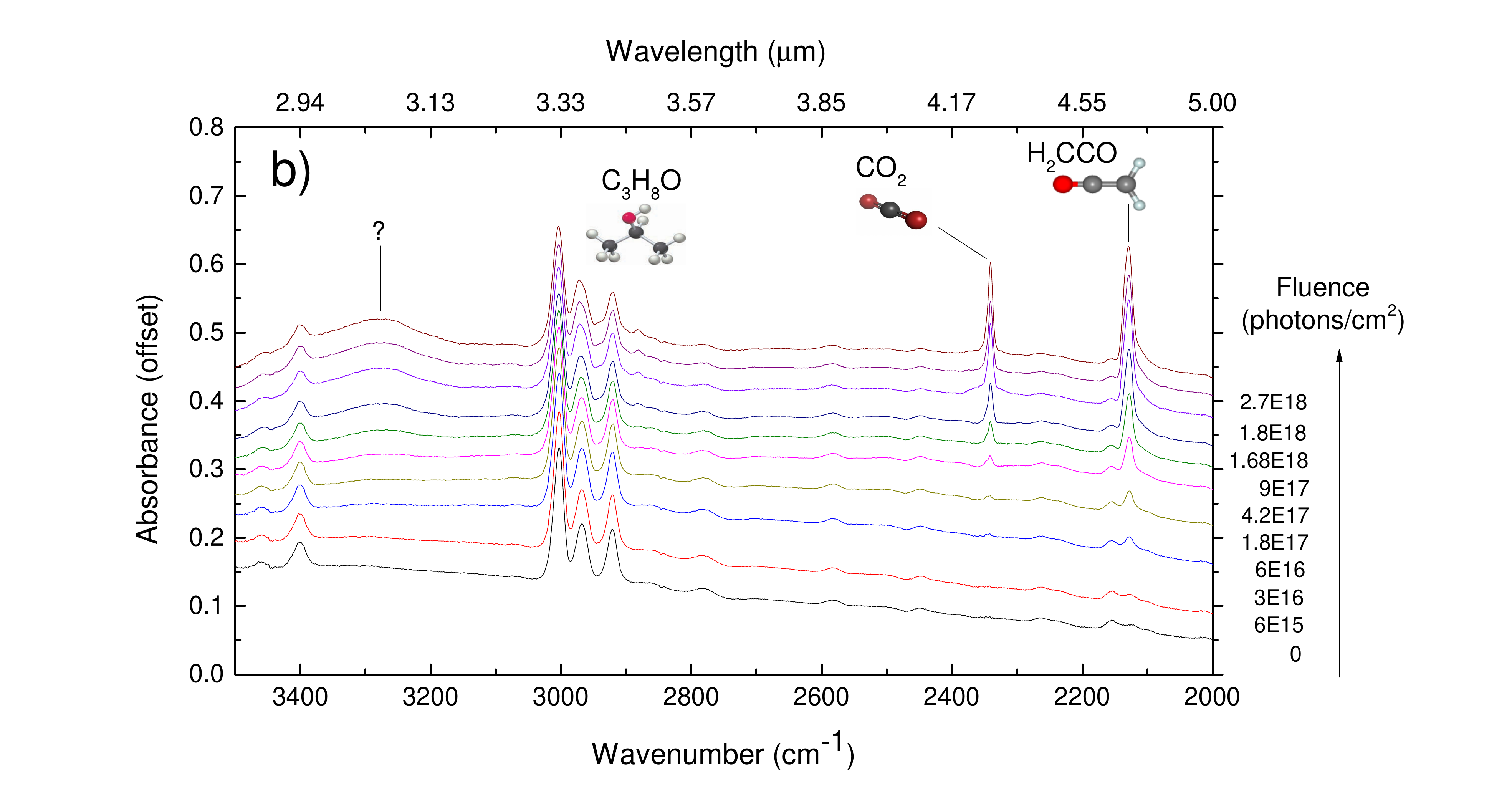}

\includegraphics[height=7cm, width=0.80\textwidth]{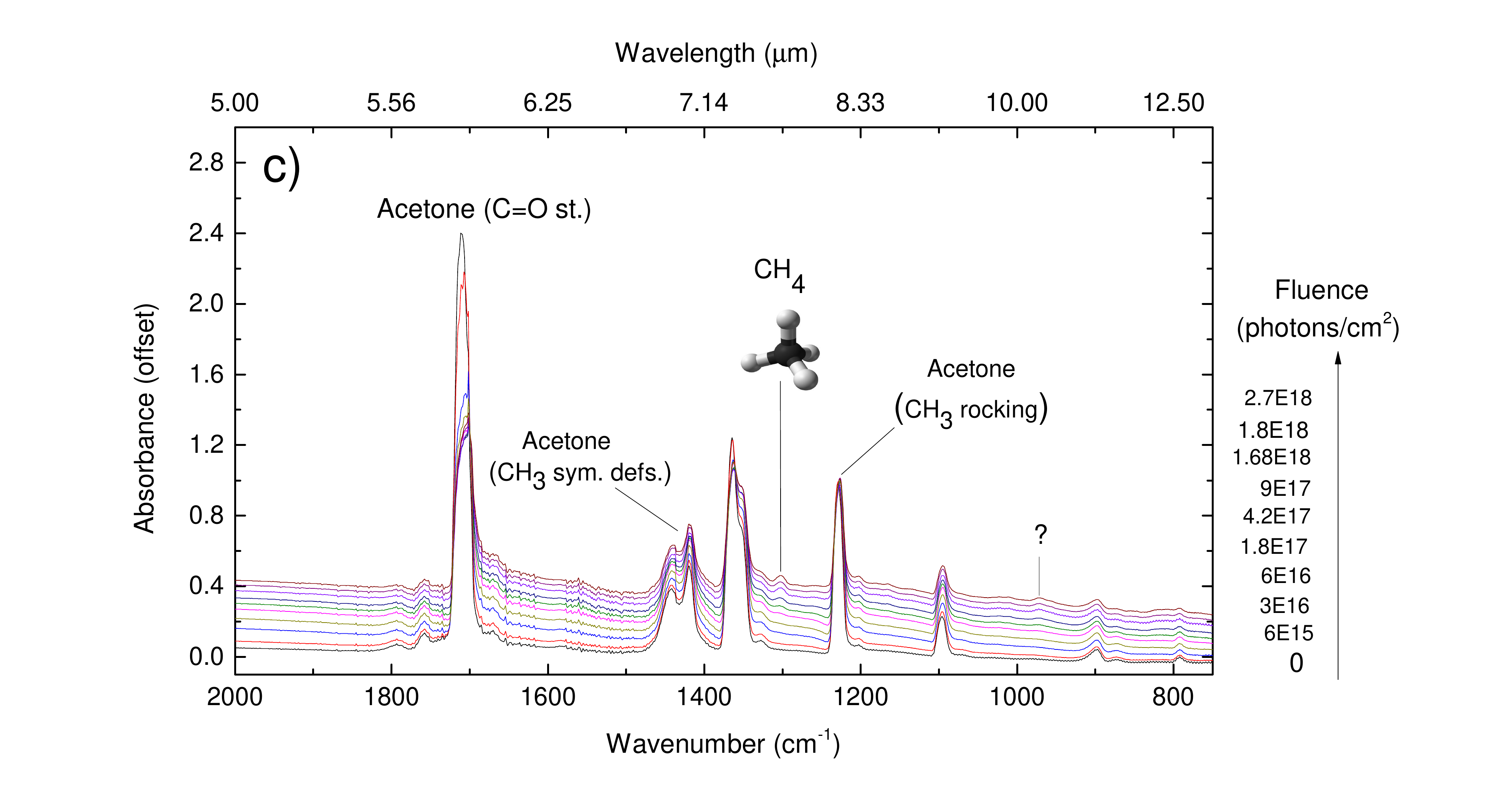}
\caption{\label{fig:1} Acetone infrared spectra. The bottom most curve correspond to the non-irradiated sample, going up fluence increases according to side indication. a) Main spectra set in the range of 4000 to 600 cm$^{-1}$. The inset plot shows the region of the father bond CO to exemplify the destruction of acetone molecules. b) Spectra in the range of 3500 to 2000 cm$^{-1}$. New peaks are identified as new molecules formed during photolysis c) Details of the infrared spectra from 2000 to 750 cm$^{-1}$, where new produced molecules are also marked. See details in the text.}
\end{figure*}

\begin{table*}
\centering
\caption{\label{tbl:example}Identified peaks in the infrared spectra of acetone ice (unirradiated). Wavenumber, wavelength, vibration mode, band strength, integration range and references.}
\begin{tabular}{lccccr}
 \hline
Wavenumber  & Wavelength  & Assignments & Band Strength  &  Region & References \\
\hline
(cm$^{-1}$) & ($\mu$m) & & ($\times 10^{-17}$ cm molecule$^{-1}$$\pm \sim5$\%) & (cm$^{-1}$) & \\
\hline
3400.2    & 2.94   & no assignment  &    -    &  3440-3380   &   -            \\
{\bf 3002.7} & 3.33 & {\bf CH$_3$ asym. stretch}  & {\bf 0.39 }  &  3033-2987   &  [1]-[4] \\
2967.3 & 3.37 & CH$_3$ asym. stretch  & - & 2986-2948   & [1]-[4] \\ 
2920.0 & 3.42 & CH$_3$ asym. stretch  & - & 2945-2896  & [1]-[4] \\
1757.3 & 5.7 & CH$_3$ asym. stretch  & - & 1771-1747  & [1]-[4] \\
{\bf 1709.7} & 5.84 & {\bf C$=$O stretch}      & {\bf 0.267}  & 1730-1689  & [1]-[4] \\
1442.0 & 6.93 & CH$_3$ sym. deformation & 0.919 & 1475-1432  & [1]-[4] \\
1419.9 & 7.04 & CH$_3$ sym. deformation & - & 1430-1394  &  [1]-[4] \\
1363.7 & 7.33 & CH$_3$ sym. deformation & 1.39  & 1383-1336  &  [1]-[4] \\
{\bf 1228.5} & 8.14 & {\bf CC asym. stretch} & {\bf 0.735}  & 1246-1211 &  [1]-[4] \\
1096.4 & 9.12 & CH$_3$ rocking mode   & 0.157 & 1110-1082  & [1]-[4] \\
898.0 & 11.13 & CH$_3$ rocking mode   & 0.0832 & 918-884 &  [1]-[4] \\
791.3 & 12.63 & CC sym. stretch   &   0.0159    & 802-778  &  [1]-[4] \\
\hline
\end{tabular}\\
{\raggedright Notes: [1]~\citet{Dellepiane1966Apr};
[2] \citet{Hudson2018Mar}; [3] \citet{Hudson2018Feb}; [4] \citet{Hudson2019May}.
}
\end{table*}

\begin{table*}
\centering
\caption{\label{tbl:exampletwo}Identified new peaks in the infrared spectra of simulated ice (irradiated sample). Wavenumber, wavelength, vibration mode, band strength, integration range and references.}
\begin{tabular}{lccccr}
 \hline
Wavenumber  & Wavelength  & Assignments & Band Strength  &  Region & References \\
\hline
(cm$^{-1}$) & ($\mu$m) & & ($\times 10^{-17}$ cm molecule$^{-1}$$\pm \sim5$\%) & (cm$^{-1}$) & \\
\hline
  3274.5 & 3.05   &   \textbf{?}    &   -  &  3369-3163   & -      \\
  2881.1 & 3.47 & C$_3$H$_8$O &    -  & 2892-2870  & [1] \\
  2340.0 & 4.27 & CO$_2$ & 11.8 & 2356-2330   & [2]\\
  2128.4 & 4.69 & H$_2$CCO(CO) & 12 & 2148-2018  & [3]\\
  1301.9 & 7.68 & CH$_4$ &  0.97  & 1313-1289  & [4]\\
  971.2 & 10.29 & {\it CH=CH bending} & - & 988-955  & [5]\\
\hline
\end{tabular}\\
{\raggedright Notes: [1]~\citet{Hudson2018Feb};
[2] \citet{Gerakines2015Jul}; [3] \citet{Berg1991Apr}; [4] \citet{Gerakines2015Jun}; [5] \citet{Matrajt2013Feb}.
}
\end{table*}

\subsection{Photolysis induced by the broadband X-ray in the frozen samples}

\begin{figure*}
\hspace*{-1.0cm}
\includegraphics[width=0.57\textwidth]{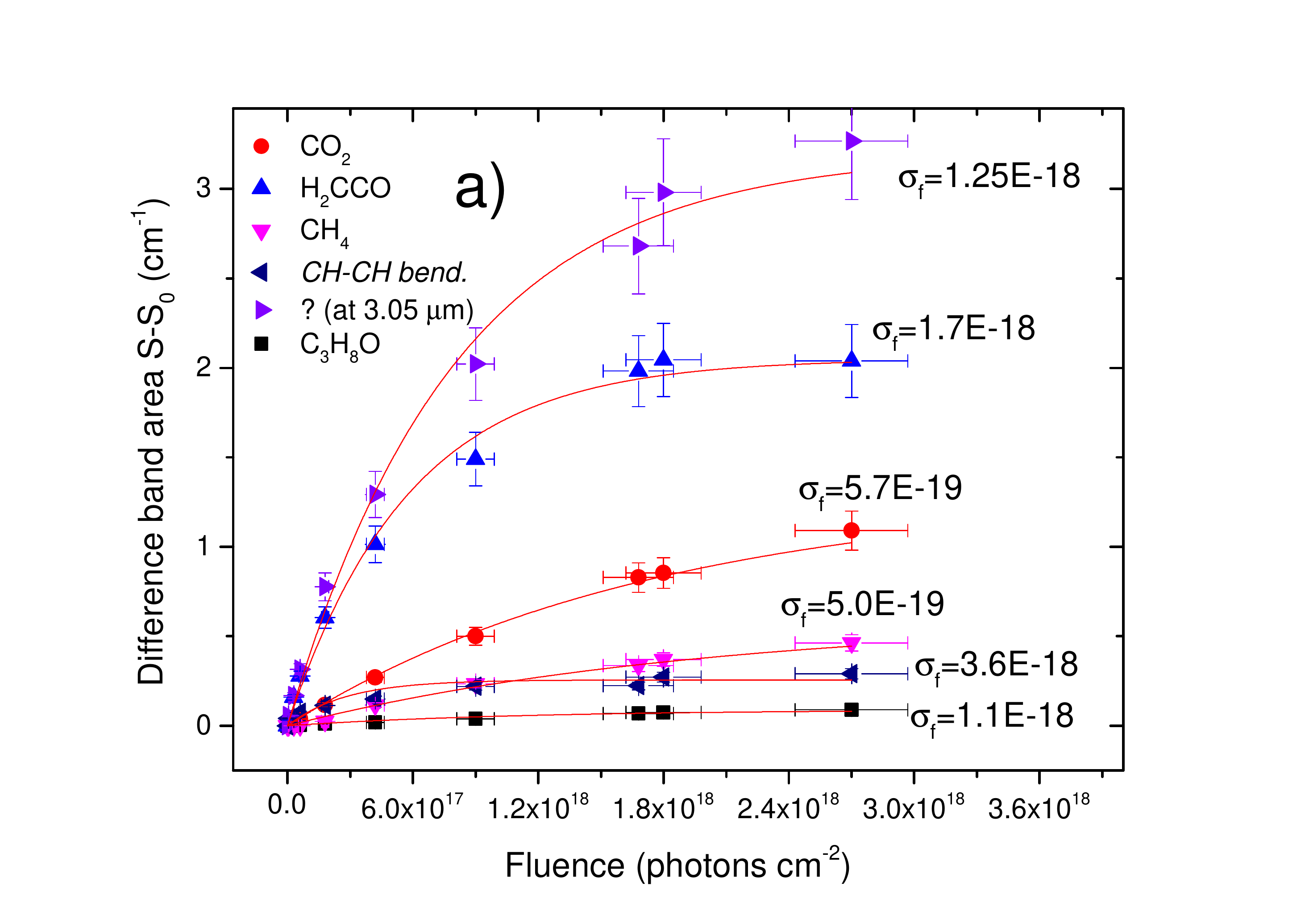} 
\hspace*{-1.8cm}
\includegraphics[width=0.57\textwidth]{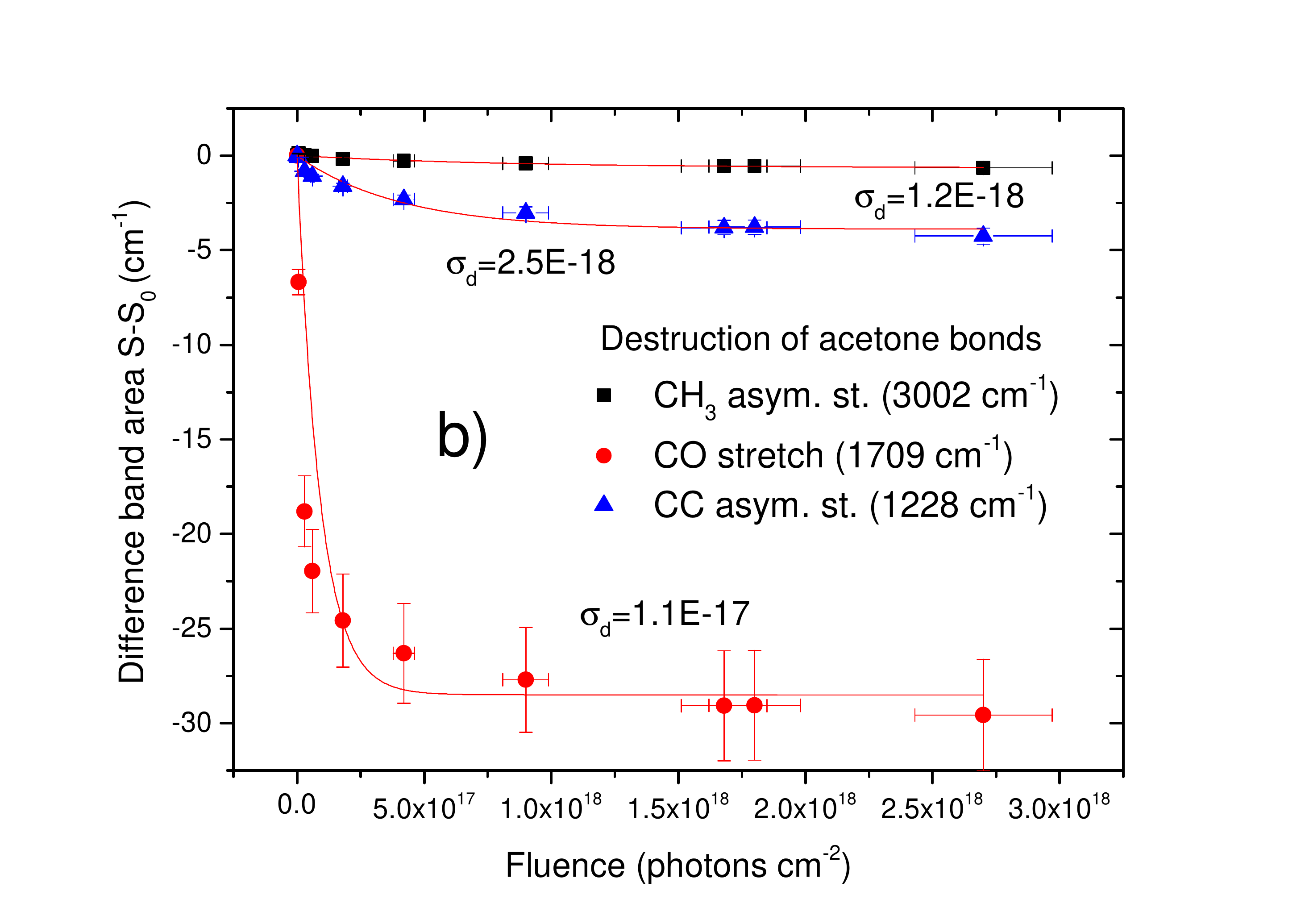}
\caption{\label{fig:2}Difference band area versus fluence. The red curves are the best fit considering the exponential expression given by equation \eqref{cs}. We determine the formation (panel a) and destruction (panel b) cross sections for the selected daughter species and father bonds, respectively.}
\end{figure*}

X-rays induce chemical changes in the ice that can be observed in Fig. \ref{fig:1}, where we show the infrared spectra of the samples during irradiation. Panel a of Fig. \ref{fig:1} shows the selected infrared spectra in the range of 4000 cm$^{-1}$ to 600 cm$^{-1}$, and the inset plot shows the same spectra in the range of 1740 to 1680 cm$^{-1}$. The inset graph demonstrates how the peak of the father bond CO decreases due increasing incoming radiation fluence. Changes on shape of the peak can be influenced by changes on molecular vicinity \citep{Pilling2020}. Fluence increases from bottom to top in all panels of Fig. \ref{fig:1}, from 0 to the maximum fluence being $2.7\times 10^{18}$ photons cm$^{-2}$. After some value of fluence the band area does not change significantly with increasing fluence. Panel b of Fig. \ref{fig:1} shows a region of the selected spectra where new peaks arise. In the panel c of Fig. \ref{fig:1} we observe the spectra in the range $2000-750$ cm$^{-1}$. This range was chosen in order to highlight new peaks that appear in this region.

Fig. \ref{fig:1} clearly shows that in increasing fluence the processes of dissociation are greater, and thus new molecular species come to light. We identify the new species by comparing the peak positions with similar works on acetone ices. New observed peaks were: a flat peak in around 3 $\mu$m that could be associated to amorphous water, but since several other alcohols have a similar peak and water is not a highly expected acetone radiation product (see section \ref{reactions}), we could not make any strong assignment to this peak. Some new peaks in the range of $2000-3500$ cm$^{-1}$, which we assign to isopropanol, ketene and carbon monoxide according to the work of \cite{Hudson2018Feb}. Other two peaks in the region $2000-600$ cm$^{-1}$ we assign to CH$_4$ and -CH=CH- bending \citep{Hudson2018Feb,Matrajt2013Feb}. The most sharp and visible peak is the CO$_2$ one. CO$_2$ could be a contaminant, however the CO$_2$ peak is not present in our control experiments, such as monitoring the ice sample before irradiation to confirm that without irradiation sample remains unaltered, or also irradiation of the bare substrate. Also, the CO$_2$ peak increases with fluence showing this is a molecule being formed due to irradiation. Due the vacuum chamber pressure to be $10^{-9}$ mbar, some residual gas could contaminate the sample, but this is accounted for our error bars estimates. The H$_2$CCO peak can be overlapped with a CO one since both are expected radiation products and have a peak near 2128 cm$^{-1}$. However, as stated in \cite{Hudson2018Feb}, the shift of this peak to higher wavenumbers with increasing fluence indicates this peak is most due to formation of ketene. In comparison with the work of \cite{Hudson2018Feb}, he observed peaks for CH$_4$, CO, H$_2$CCO and C$_3$H$_8$O. Here, we observed and confirm a peak for CO$_2$ as well other similar peaks presented in the work of \cite{Hudson2018Feb} for isopropanol and ketene. Photo-irradiation of pure acetone ice with X-rays was also performed in \cite{Almeida2014Mar}, where authors employed a photo stimulated desorption (PSID) and took the positive PSID spectra to analyze acetone fragmentation. They have thus detected lots of cations as fragments, such as HCO$^+$, CO$^+$, H$^+$, C$_2$CH$_3$$^+$, CO$_2$$^+$, H$_2$CO$^+$, etc.. Tables \ref{tbl:example} and \ref{tbl:exampletwo} present a compilation of peak positions (in wavenumbers and wavelength), our corresponding assignments, band strengths and references used to determine molecular assignments. Question mark in Table \ref{tbl:exampletwo} is a peak probably related to several different molecules being formed in the ice as explained before. Note that ketene band strength was taken from \cite{Berg1991Apr} and it was also used as reference value in other recent works \citep{Hudson2013Jul,Bergner2019Mar}. No reference band strength for C$_3$H$_8$O were found.

\subsection{Formation and destruction cross-section}

The formation and destruction cross-sections can be determined by monitoring the numerical evolution of the molecular band area with the use of the following equation
\begin{equation}\label{cs}
S-S_0= S_\infty \times \left(1-e^{-\sigma_{d,f}\times F}\right) ~ [\rm cm^{-1}],
\end{equation}
where $S$, $S_0$ and $S_\infty$ are the selected areas of the infrared band related to a specific molecular vibration mode at a given fluence, at beginning of the experiment (i.e., zero fluence for the unirradiated sample) and at highest fluence, respectively. $F$ is the fluence in the units of cm$^{-2}$, and $\sigma_{d;f}$ represents formation and destruction cross-sections in units of cm$^{2}$ (depending on the case, i.e., $\sigma_f$ is the cross-section for the new species in the ice – daughters – and $\sigma_d$ describes the destruction cross-section of the infrared bands of acetone). This procedure presents some advantages in comparison with the methodology employed where authors use column density since it does not introduce errors about band strengths as it takes into account only the band area of selected peaks in the IR spectra.

Fig. \ref{fig:2} presents the difference band area ($S-S_0$) of selected peak in the infrared spectra to quantify the formation of new species (panel a) and destruction of specific bonds of frozen acetone (panel b) as a function of radiation fluence (soft X-rays in the range $6-2000$ eV). To determine the molecular dissociation of acetone due to X-ray bombardment (sensitivity to radiation) we considered three molecular bonds (-CH$_3$ asym. str., -C=O str., and –C-C asym. str.) as are marked with bold in Table \ref{tbl:example}. The most sensitive bond to radiation was the -C=O, with the larger destruction cross-section ($\sigma_d = 1.1 \times 10^{-17}$ cm$^{2}$), this is expected since the cation (CH$_3$COHCH$_3)^+$ would be one of the first acetone radiation products to be formed \cite{Hudson2018Feb}. As discussed in details in several previous papers \citep[see, e.g.][]{Andrade2013Apr,Portugal2014Jul,Freitas2020}, different bonds have different sensitivity to radiation-induced processes. This indicates that the effective cross section calculated by such bonds depends on the considered bond. As discussed by \cite{Portugal2014Jul}, it is possible to employ an average value as well as take into account a specific value (e.g. vibration mode of a molecular backbone) as representative for the destruction of the molecule in the ice phase. Here, we considered the C-C vibration of acetone as representative for the molecule itself. The reader must remember that, here, the effective destruction cross section (in solid phase) determined in this manuscript takes into account a large number of molecular scenario (locally) being exposed and destructed by radiation within the sample.

To calculate formation cross-sections, we take into account the four daughter molecules CO$_2$, H$_2$CCO, CH$_4$ and C$_3$H$_8$O. We also calculated effective formation cross-section for the peaks 3274 and 971 cm$^{-1}$. As discussed in some previous works \citep[see][]{Almeida2014Mar,deA.Vasconcelos2017Nov} the destruction cross section of frozen molecules by the bombardment of either heavy ions or energetic photons can be modeled as the following power-law
\begin{equation}\label{sigma}
  \sigma_d=a\times E^n \quad {\rm [cm^2]},
\end{equation}
where $n$ determines the power-law order, $\sigma_d$ is the molecular dissociation cross-section in cm$^2$, $E$ is the projectile energy in MeV and $a$ is a proportionality constant. 

\begin{figure}
\hspace*{-1.2cm}
 \includegraphics[height=7.8cm]{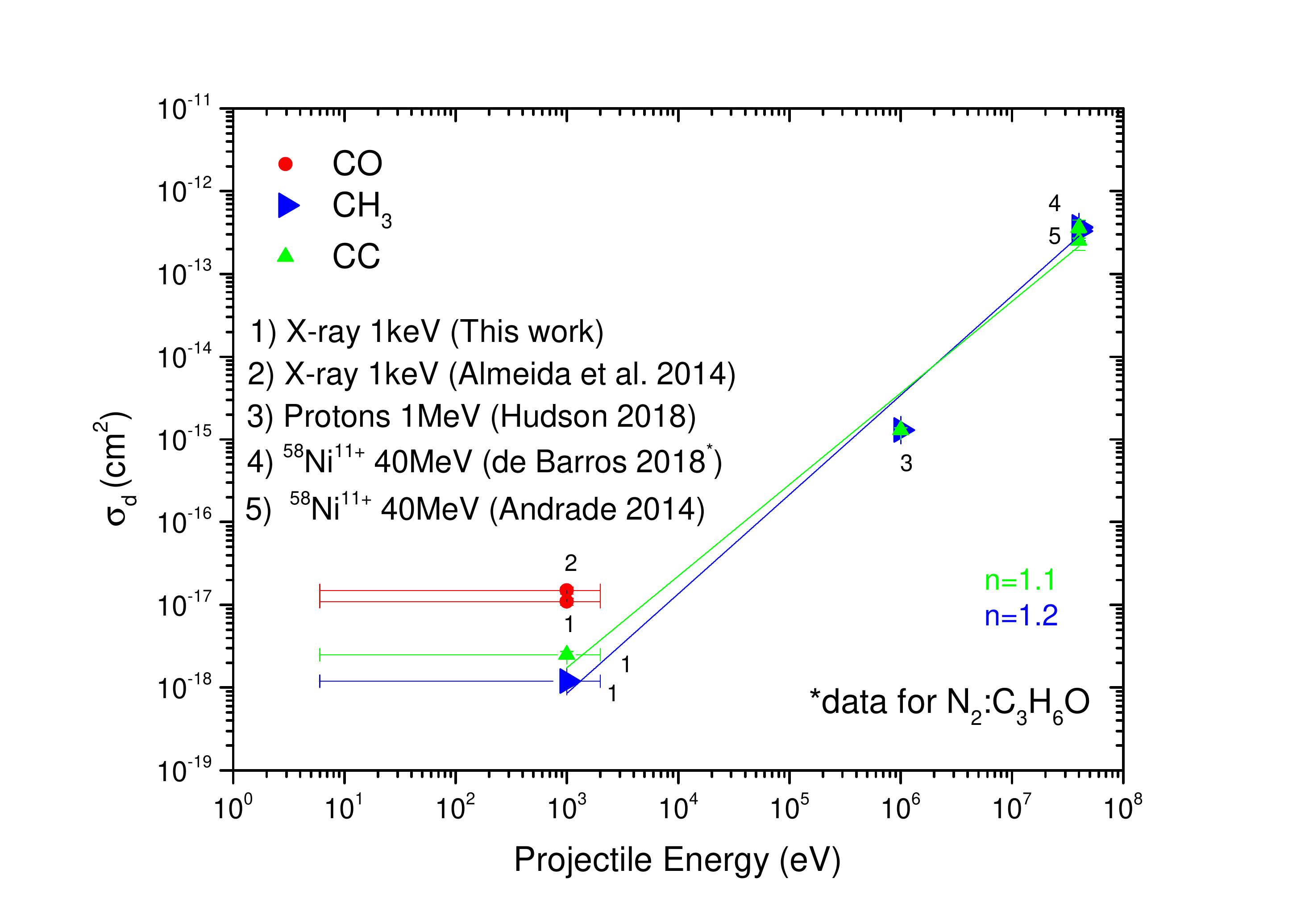}
 \caption{\label{sigmaXenergy} Dissociation cross-section as a function of projectile energy. The lines corresponds to the best fit employing equation \eqref{sigma} and considered only CH$_3$ and CC vibration modes.}
\end{figure}

In Fig. \ref{sigmaXenergy} we show the destruction cross-sections calculated for several types of sources of ionizing radiation from heavy ions to X-rays. The molecular dissociation cross-section obtained in this work is in good agreement with the work of \cite{Almeida2014Mar}. In particular, the value of destruction cross-section of the CO bond in our work is $1.1 \times 10^{-17}$ cm$^{2}$, while in the work of \cite{Almeida2014Mar} the destruction cross-section for CO was $1.5 \times 10^{-17}$ cm$^{2}$, showing that our obtained cross section values are feasible.

The effective destruction and formation cross-sections are listed in Table \ref{tbl:examplethree}. The greater formation cross-sections of identified molecules were determined for H$_2$CCO and C$_3$H$_8$O, which is consistent with reaction routes proposed by \cite{Hudson2018Feb}.

\begin{table}
\centering
\caption{Obtained destruction and formation cross-sections for father bonds and daughter species in the ice sample.\label{tbl:examplethree}}
\begin{tabular}{cc}
\hline
Molecule (bond) & Cross-section $\sigma_{d,f}$ ($10^{-18}$cm$^{2}$)\\
\hline
Acetone (CH$_3$) & $1.2$ \\
Acetone (CO) & 11 \\
Acetone (CC$_2$) & 2.5 \\
? (at $3.05~\mu$m) & 1.25 \\
C$_3$H$_8$O & 1.1\\
H$_2$CCO & 1.7 \\
CO$_2$ & 0.57 \\
CH$_4$ & 0.5\\
{\it CH=CH bending} & 3.6\\
\hline
\end{tabular}
\end{table}

\subsection{Chemical equilibrium}

\begin{figure*}
 \includegraphics[height=10cm]{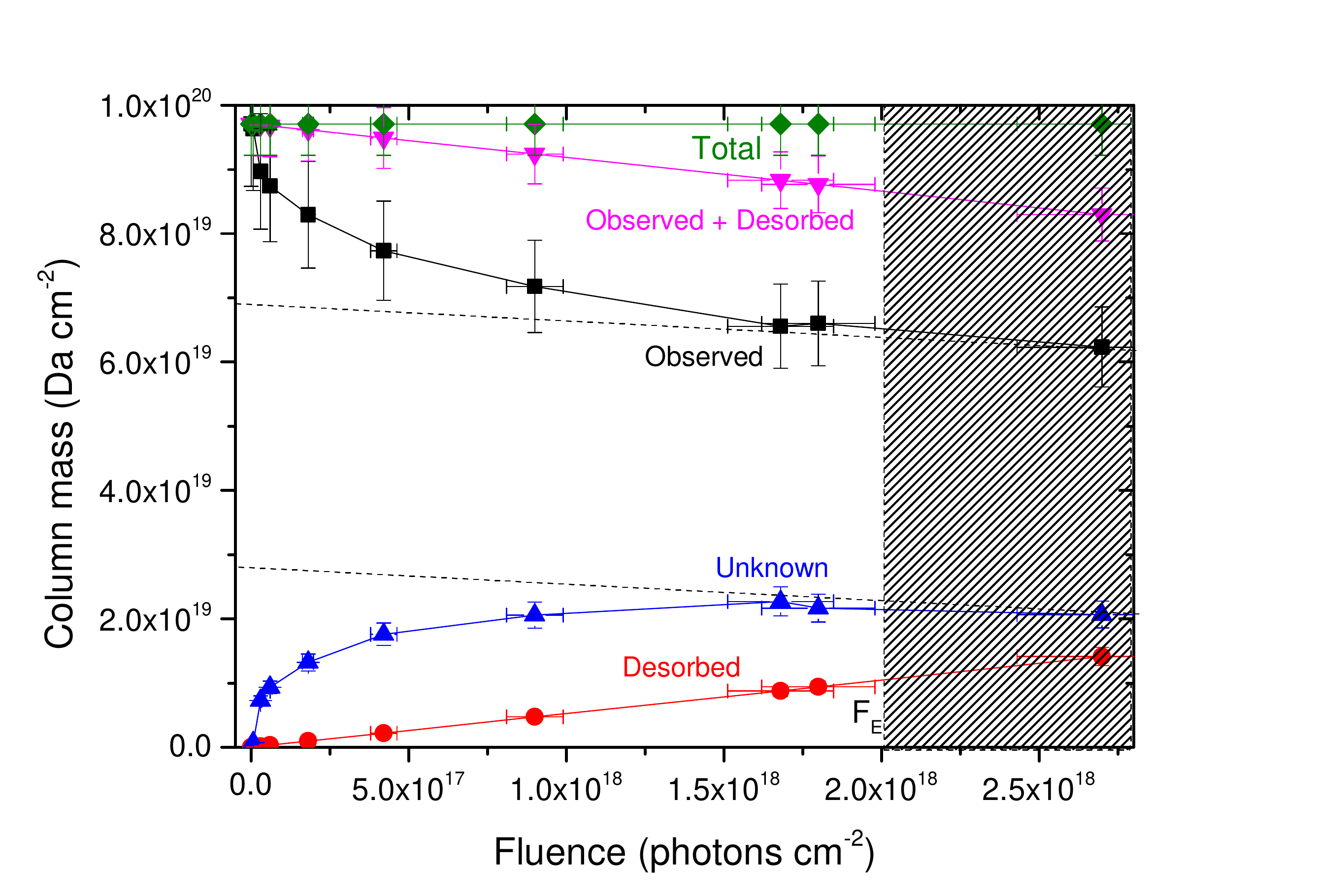}
 \caption{\label{colmassXfluence} Calculated column mass of molecular species in the acetone ice as a function of fluence using infrared spectroscopy and the procedure described in equations \eqref{2}-\eqref{4}. Chemical equilibrium is reached for large fluences where unknown and observed column masses have the same linear slope. Dashed lines are employed for better visualization. The desorbed mass, $M_{\rm DES}$, is determined using equation \eqref{3}, and the fitted sputtering rate is $Y=0.09$ molecules photon$^{-1}$. The green points and lines represent the sum of all column masses and it can be observed that it is approximately a constant function. The shaded region corresponds to chemical equilibrium.}
\end{figure*}

\begin{figure*}
 \includegraphics[height=10cm]{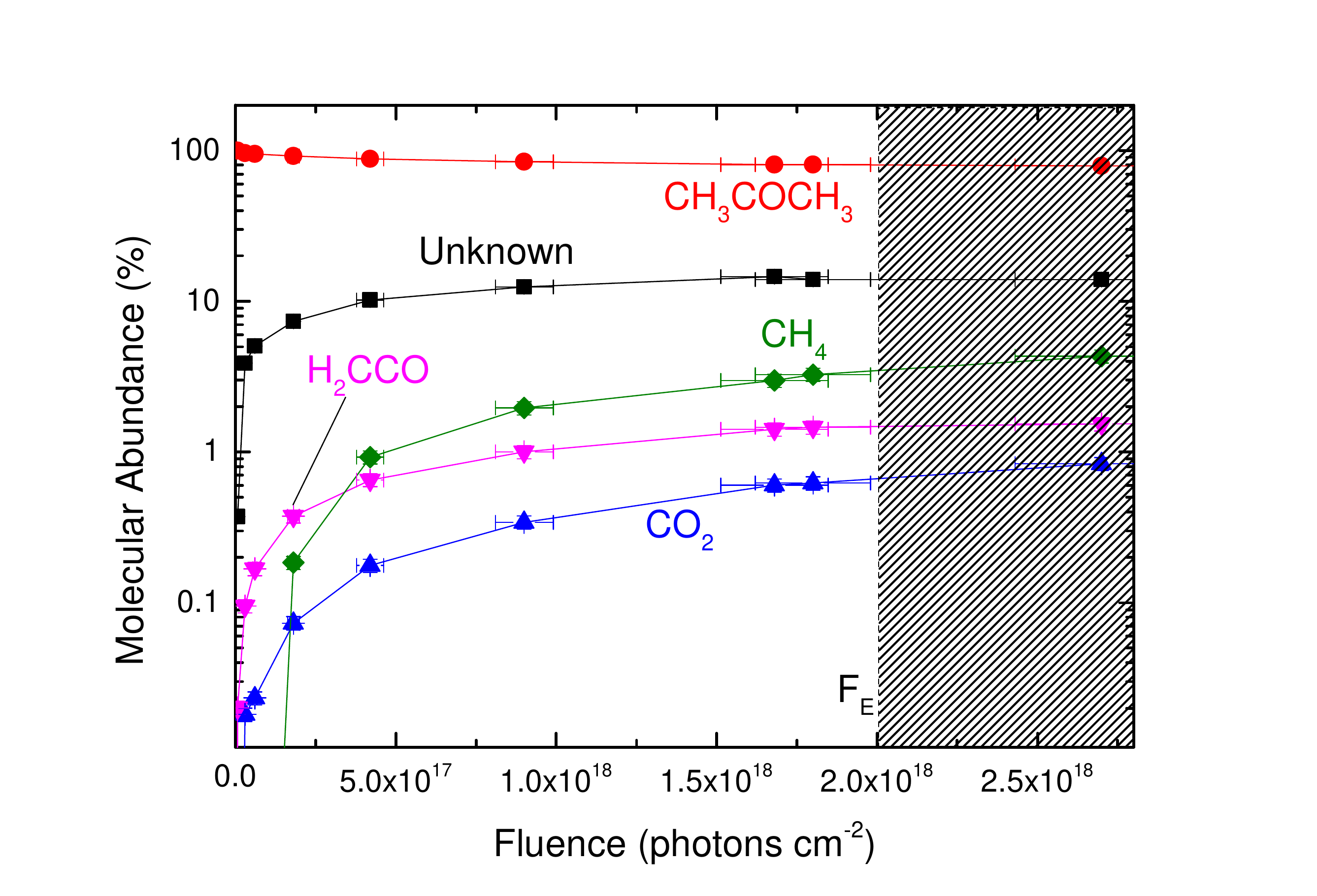}
 \caption{\label{molecular_abundance} Molecular abundance (in percentage) as a function of fluence during irradiation of acetone ice at 12 K by soft X-rays. The estimated value for unknown species are also showed. Shaded region correspond to chemical equilibrium.}
\end{figure*}

From the experiments, we observe that at large fluences the band-areas and its related column mass and column densities of parent and daughter species tend asymptotically to a horizontal plateau. This implies a chemical
equilibrium scenario in the ice, which means that parental and daughter species are dissociated/produced at a similar rate. As discussed by \cite{Pilling2019} the chemical equilibrium can be defined in a different manner as the moment when the summed chemical abundances does not change considerably. It is interesting to note that as the band-area tends to a plateau after some value of fluence it indicates that larger fluences yield the same chemical equilibrium, since it induces the same reaction rates for parent and daughter species in the ice. Such behavior is also observed in the irradiation of other ices by different ionizing sources \citep[e.g.][]{Pilling2019,Pilling2010Jan,deA.Vasconcelos2017Nov,Rachid2017Dec,Freitas2020}.

Due to continuous desorption induced by the X-ray irradiation the amount of molecules present in the samples reduces slowly even after reaching chemical equilibrium. In a previous recent work \citep{Pilling2019}, authors elaborated a method to determine the amount of desorbed species to gas-phase due X-ray irradiation of the sample using the above concept of chemical equilibrium. The authors have also determined how to calculate the percentage of molecules in the ice (chemical abundances) by defining also how to calculate the mass of unknown species. Molecules of unknown species are due to several reasons, such as: i) vibration modes related with a vanishing dynamic dipole moment or homo-nuclear diatomic molecules (no permanent dipole), ii) abundances below the spectrometer detection limit, iii) infrared bands overlapped with other bands and iv) Infrared bands mixed with noise in the spectra and therefore being difficult to be measured/quantified.
To get a reasonable estimate of the amount of unknown species in the sample and also of those (known and unknown) desorbed from the sample surface it is necessary to determine the evolution of the column masses of different species in the ices as a function of fluence. First, to calculate the molecular column density we use the following equation \citep{Pilling2010Jan,deA.Vasconcelos2017Nov}
\begin{equation}
N=2.3 \frac{S}{A} ~~ {\rm [molecules ~ cm^{-2}]},
\end{equation}
where $S$ is the integrated absorbance in units of cm$^{-1}$, i.e., it corresponds to the band-area in cm$^{-1}$ and $A$ is the band strength in units of cm molecule$^{-1}$. This procedure allows us to calculate the column densities of the father and daughter molecular species. Given the column densities, we follow the same procedure applied in \citep{Pilling2019,deA.Vasconcelos2017Nov,Pilling2010Jan} to estimate the column masses of the father, daughter, desorbed and unknown species. Some hypothesis are taken into account, as we will review it below. 

The column mass of a given specie $i$, in units of Da cm$^{-2}$, as a function of fluence is
\begin{equation}\label{2}
  M_i(F)= N_i(F) \times \overline{MM_i} ~ [{\rm Da ~cm^{-2}}],
\end{equation}
where $N_i$ is the column density of the $i$-specie at a given fluence $F$ and $\overline{MM_i}$ is the average molecular mass of a given molecule $i$, in units of Da per molecule. The first hypothesis is that the estimated column mass of all desorbed species (father, daughter and unknown) from the ice sample during X-ray bombardment can be a function of fluence as approximately
\begin{equation}\label{3}
 M_{\rm DES} \approx \overline{MM_0} \times Y \times F ~~{\rm [Da~ cm^{-2}]},
\end{equation}
where $\overline{MM_0}$ is the average molecular mass of the parent species, which means that molecular desorption is mainly governed by parent desorption. In addition, $Y$ represents the sputtering rate in molecules per photon.

The next step in the procedure is to determine the column mass of unknown species by invoking the mass conservation law which gives us the column mass of unknown species as 
\begin{equation}\label{4}
  M_{\rm UN}(F)= M_j(F=0) - \left( M_j (F) + M_{\rm DES}(F)\right)~~ {\rm [Da ~cm^{-2}]},
\end{equation}
where $ M_j (F)$ is the column mass of acetone for the fluence $F$ and $M_j(F=0)$ is the acetone column density at the beginning of the experiment, i.e., when the sample were not irradiated yet. $M_{\rm DES}$ is calculated using the proposed equation \eqref{3}. 

As described by \cite{Pilling2019}, the column densities for the unknown species can be quantified using the following hypothesis 
\begin{equation}
  N_{\rm UN}= \frac{M_{\rm UN}}{\overline{MM_0}} ~~{\rm [Da ~per~ cm^2]}.
\end{equation}

To determine mathematically the chemical equilibrium we calculate the following quantity
\begin{equation}\label{MA}
  \Delta MA= \sum_i |MA_i(F_k)-MA_i(F_{k-1})|,
\end{equation}
where $MA_i$ represents the molecular abundance of the $i$-specie. Thus, chemical equilibrium is reached for a given fluence $F_k$ which provides $\Delta MA<1\%$. Employing equation \eqref{MA} we obtain the fluence of chemical equilibrium as approximately $2\times 10^{18}$ photons per cm$^2$.

In Fig. \ref{colmassXfluence} we present the column mass evolution with fluence of all observed, unknown and desorbed molecules. For the sake of clearness we compute observed plus unknown column masses and the sum of observed, unknown and desorbed column masses. The sum of column masses as expected by the mass conservation law is a constant function. To determine the sputtering yield we start the calculations using $Y=0.2$, posteriorly, this value was fitted using an iterative method in order to obtain the same linear slope for unknown and observed column masses when the chemical equilibrium is reached. The best value of the sputtering yield has shown to be approximately $Y=0.09$. In the previous work of Pilling and collaborators for water-rich ices \citep{Pilling2019} the values of sputtering rate were shown to have approximately the same magnitude of the one obtained here.

The molecular abundance in percentage as a function of X-ray fluence for the studied ice at 12 K is presented in Fig. \ref{molecular_abundance}. The hatched region indicates the region in which chemical equilibrium happens. The fluence in which chemical equilibrium is reached (called equilibrium fluence, $F_E$) is indicated near x-axis. The value determined in this work - $F_E \sim 2 \times 10^{18}$ photons per cm$^2$ - is closer to  previous experiments employing X-rays in astrophysical ice analogs \citep[e.g.][]{Pilling2015Sep,deA.Vasconcelos2017Nov,Rachid2017Dec,Pilling2019}. Table \ref{tbl:molec_abun} presents the molecular abundances of acetone, identified irradiation products and unknown species. From Table \ref{tbl:molec_abun} we observe that the unknown component of the ice is abundant and although its composition to be unrecognized we can speculate about it. C$_3$H$_8$O is a reaction product with a identified peak from the ice spectra, so it is a molecule present in the ice, however, without reliable band strength we could not account for it on molecular abundances, so this molecule contributes to the unknown component. Other molecules such as C$_2$H$_2$, C$_2$H$_4$, C$_2$H$_6$, C$_2$H$_4$O, CO, CH$_2$O and CH$_2$CCH$_2$ are all molecules expected to be radiation products \citep[see][]{Hudson2018Feb} and could thus be present as components of the unknown section, however to state their abundances is not possible, we can only state they were not detected and may probably contribute to non-assigned peaks. 

\begin{table}
\centering
\caption{\label{tbl:molec_abun}Molecular abundances at chemical equilibrium or Equilibrium Branching Ratio (EBR\%) for the acetone ice. Acetone molecular abundance at the beginning of the experiment is assumed to be 100\%.}
\centering
\begin{tabular}{cc}
\hline
Molecule & EBR(\%) \\
\hline
Acetone (CH$_3$COCH$_3$) & $79.1$ \\
H$_2$CCO & 1.5 \\
CO$_2$ & 0.8 \\
CH$_4$ & 4.3 \\
Unknown & 14.3\\
\hline
\end{tabular}
\end{table}

\section{Astrophysical Implications}\label{sectIV}

One interesting feature of ionizing X-ray radiation is that it is produced in several astrophysical environments, such as young stellar objects, main-sequence stars, and compact stars including white dwarfs, neutron stars and black holes. The X-ray generated by those objects interact with solid and ice bodies/grains and gas in their vicinities, triggering photochemical processes (e.g. breaking chemical bonds, radical production, molecular formation and desorption processes) in regions such as dense molecular clouds, dust disks and rings, planetary debris and astrophysical ices. 

\subsection{X-rays estimate}

To extrapolate laboratory conditions to some space environments, we estimate the X-rays flux (from 6 to 2000 eV) in the vicinity of selected sources as some young stars, main-sequence stars (sun and OB stars), and around compact objects (white dwarfs, neutron stars and black holes) by considering literature data \citep{Pilling2019,deA.Vasconcelos2017Nov}. Those photon fluxes are employed together with current experimental data to calculate the timescale of ices (typical acetone-rich ices) to reach chemical equilibrium in a given astrophysical environment.

The estimated X-ray fluxes at selected distances employed in this work and respective timescales for ices to reach chemical equilibrium at such distances are listed in Table \ref{photonfluxes}.

\begin{table*}
\centering
\caption{\label{photonfluxes}Estimated X-ray photon flux for selected space environments and its respective timescale to reach chemical equilibrium.}
\begin{tabular}{lccr}
\hline \hline
Space environment & Estimated photon flux & $TS_E$ & References \\
\hline
 & (cm$^{-2}$s$^{-1}$) & (year) &  \\
\hline
  KBOs typical orbit ($\sim 40$ AU) & $5\times 10^3$ & $1.2\times 10^7$ & [1,2]\\
  Triton orbit ($\sim 40$ AU) & $9\times 10^6$ & $6.6\times 10^3$ & [1,2]\\
  Saturn orbit ($\sim 9.5$ AU) & $8\times 10^9$ & $7.5$ & [1,2]\\
  Earth orbit (Sun's typical flux at$\sim 1$ AU) & $3\times 10^{13}$ & $2\times 10^{-3}$ & [2]\\
  YSO model 1 (flux at 30 AU) & $1\times 10^6$ & $6\times 10^{4}$ & [3]\\
  YSO model 2 (TW Hydra at 40 AU) & $9\times 10^{11}$ & $6.6\times 10^{-2}$ & [4]\\
  YSO model 3 (typical flux at 1 AU) & $6\times 10^{15}$ & $1\times 10^{-5}$ & [5]\\
  Laboratory & $1\times 10^{14}$ & $6\times 10^{-4}$ & [1]\\
  White Dwarf (WD 0736+053 flux at 1 AU) & $4\times 10^{8}$ & $1.5\times 10^{2}$ & [6]-[11]\\
  White Dwarf (WD 0736+053 flux at 1 LY ($6.3\times 10^4$ AU)) & $1\times 10^{-1}$ & $6\times 10^{11}$ & [6]-[11]\\
  White Dwarf (WD 1314+293 flux at 1 AU) & $2\times 10^{9}$ & $30$ & [6]-[11]\\
  White Dwarf (WD 1314+293 flux at 1 LY ($6.3\times 10^4$ AU)) & $6\times 10^{-1}$ & $1\times 10^{11}$ & [6]-[11]\\
  White Dwarf (Sirius B flux at 1 AU) & $1\times 10^{10}$ & $6$ & [9]\\
  White Dwarf (Sirius B flux at 1 LY ($6.3\times 10^4$ AU)) & $4$ & $1.5\times 10^{10}$ & [9]\\
  White Dwarf (WD 0216-032 flux at 1 AU) & $2\times 10^{11}$ & $3\times 10^{-1}$ & [6]-[11]\\
  White Dwarf (WD 0216-032 flux at 1 LY ($6.3\times 10^4$ AU)) & $44$ & $1.4\times 10^{9}$ & [6]-[11]\\
  Type B Star (CD-38 11636(O8) flux at 1 AU) & $5\times 10^{12}$ & $1.2\times 10^{-2}$ & [12]\\
  Type O Star (HD 92644 flux at 1 AU) & $1\times 10^{13}$ & $6\times 10^{-3}$ & [12]\\
  Neutron Star binary (SAX J1750.8-2900 at 1 AU) & $7\times 10^{14}$ & $8.6\times 10^{-5}$ & [13]-[15]\\
  Neutron Star binary (SAX J1750.8-2900 at 2.25 LY ($1.4\times 10^5$ AU)) & $4\times 10^{4}$ & $1.5\times 10^{6}$ & [13]-[15]\\
  Vela Pulsar (PSR B0833-45 flux at 1 AU) & $1\times 10^{14}$ & $6\times 10^{-4}$ & [16]\\
  Vela Pulsar (PSR B0833-45 flux at 2.25 LY ($1.4\times 10^5$ AU)) & $5\times 10^{3}$ & $1.2\times 10^{7}$ & [16]-[18]\\
  Crab Pulsar (PSR B0531+21 flux at 1 AU) & $2\times 10^{18}$ & $3\times 10^{-8}$ & [16]-[18]\\
  Crab Pulsar (PSR B0531+21 flux at 2.25 LY ($1.4\times 10^5$ AU)) & $1\times 10^{8}$ & $600$ & [16]-[18]\\
  Black Hole binary (A0620-00 flux at 1 AU) & $7\times 10^{10}$ & $8.6\times 10^{-1}$ & [16]\\
  Black Hole binary (A0620-00 flux at 3 LY ($1.9\times 10^5$ AU)) & $2$ & $3\times 10^{10}$ & [16]\\
  Supermassive Black Hole (Sag A* flux at 1 AU) & $2\times 10^{14}$ & $3\times 10^{-4}$ & [19]-[21]\\
  Supermassive Black Hole (Sag A* flux at 3 LY ($1.9\times 10^5$ AU)) & $6\times 10^{3}$ & $1\times 10^{7}$ & [19]-[21]\\
\hline
\end{tabular}\\
{\raggedright Notes: [1] \citet{Pilling2015Sep}; [2] \citet{Gueymard2004Apr}; [3] \citet{Walsh2012Feb}; [4] \citet{Fantuzzi2011Nov}; [5] \citet{Siebenmorgen2010Feb}; [6] \citet{O'Dwyer2003Apr}; [7] \citet{Chu2004Jan}; [8] \citet{Bilnkova2010Oct}; [9] \citet{Fleming1993}; [10] \citet{Heise1985Feb}; [11] \citet{Weisskopf2007Mar}; [12] \citet{Binder2019Apr}; [13] \citet{Wijnands2013Jul}; [14] \citet{Parikh2017Aug}; [15] \citet{vanDoesburgh2019Oct}; [16] \citet{Lewin2006Apr}; [17] \citet{vanDoesburgh2019Oct}; [18] \citet{Potekhin2019Sep}; [19] \citet{Barriere2014Apr}; [20] \citet{Marrone2008Jul}; [21] \citet{Quataert2002Aug}.
}
\end{table*}

\subsection{Main-sequence and pre-main-sequence stars}

For young stellar objects (YSOs) and for our Sun, the estimates for the photon flux were taken from \citep{Pilling2015Sep,deA.Vasconcelos2017Nov,Pilling2019}. As we can observe in Table \ref{photonfluxes}, at the distance of  30 AU of a typical YSO (YSO model 1) the timescale for ices to reach chemical equilibrium is around $10^4$ years, which is a similar value to the one obtained by \cite{Pilling2019} also employing X-rays in a mixed ice analog. In particular, the Sun's photon flux (ultraviolet and mainly X-rays in the range $6-2000$ eV) provides timescales of $\sim 10^{-3}$ year at 1 AU (Earth orbit) and 7.5 years at 9.5 AU (Saturn orbit).

The hot and massive type O and B stars (with mass greater than 8 solar masses) release large quantities of thermal radiation in UV and X-rays. The non-thermal emission arises in O/B stars from relativistic particles that are accelerated in a magnetic field or in colliding winds in regions between the stars of a binary system. Type O and B stars considered in this work have soft X-ray spectra and have high X-ray luminosity in the range of 0.5 - 7 keV. This spectrum interval is consistent with the experimental one employed in this work (6 eV - 2 keV), thus we take the luminosity to calculate the photon flux at any distance from the stars using the following relation
\begin{equation}\label{phifox}
  \phi=\frac{1}{4\pi d^2}\frac{L_x}{E} ~~{\rm [photons~ cm^{-2}~s^{-1}]}
\end{equation}
where $d$ represents the distance in cm, $L_x$ is the X-ray luminosity in units of eV s$^{-1}$ (within the selected energy range), and $E$ is the photon energy in units of eV (in this work being 1 keV taken as a fiducial value).

Here, we consider a type O star (with stellar counterpart CD-38 11636) and a type B star (whose stellar counterpart is HD 92644). The estimated photon fluxes for these stars at distance of 1 AU and its respective timescales of chemical equilibrium are presented in Table \ref{photonfluxes}. Selected O/B stars are binary stars and emission is mainly due to a colliding wind in a region between the two objects. The colliding wind is responsible for the non-thermal emission in O/B binaries. Non-thermal emission in O/B stars is highly expected to occur also in other wavelengths, such as radio. So, emission mechanisms in O/B binaries are different than that of our Sun, and part of the emission power in O/B binaries is processed by the particle wind. In our Solar System the expected distance where temperature is low enough for volatile compounds to condense into solid ice grains is around 5 AU. This is called the snow line distance of our Solar system and it corresponds to where water ice can be stable. For this particular distance the timescale for acetone ice to reach chemical equilibrium is around 18 days considering the photon flux generated by the Sun and the timescales that are 3.6 and 1.8 months for the considered type B and O stars, respectively. YSOs have photon fluxes that can be larger than the Sun's one. We considered here three models for the X-ray emission of YSOs that are described in more details in \cite{Pilling2015Sep}. Models provide X-ray fluxes listed in Table \ref{photonfluxes}, which shows also the corresponding timescale to reach chemical equilibrium in such situations. Calculations have shown that timescales to reach chemical equilibrium are far below the average lifetime of a YSO ($10^5-10^6$ years). Interstellar ice particles close to YSOs are the photo-chemical reaction pots where complex organic molecules are produced consequent to stellar radiation. Since acetone transitions were also detected in disk-like structures of protostars \citep{Isokoski2013Jun}, the present study helps us to understand the role of X-rays on photochemical processes occurring in protostar envelopes.

\subsection{Compact stars}

Compact stars (white dwarfs and neutron stars) are highly efficient radiation emitters. They can power high X-ray luminosity in the range $10^{25} - 10^{45}$ erg s$^{-1}$. Below we will approach the X-ray emission of compact stars and how this emission could be related to chemical equilibrium of ices close to the vicinities of those compact objects. It is worth to cite that we estimate the X-ray photon flux of the compact objects from its X-ray luminosity. In general, the X-ray luminosity of white dwarfs and thermal X-ray radiation of isolated neutron stars have a spectrum that ranges from 0.5 to 2 keV and from 10 eV to 3 keV, respectively. Those soft X-ray components are fairly between the interval of our experimental ionizing radiation beam (6 eV - 2 keV). Thus we have estimated photon fluxes from its X-ray luminosity by using the same procedure of equation \eqref{phifox}, i.e., we use full X-ray luminosity obtained from \citep{O'Dwyer2003Apr,Chu2004Jan,Bilnkova2010Oct}.

\begin{figure*}
\centering
\hspace*{-0.8cm}\vspace*{-0.9cm}
 \includegraphics[height=7.8cm, width=0.78\textwidth]{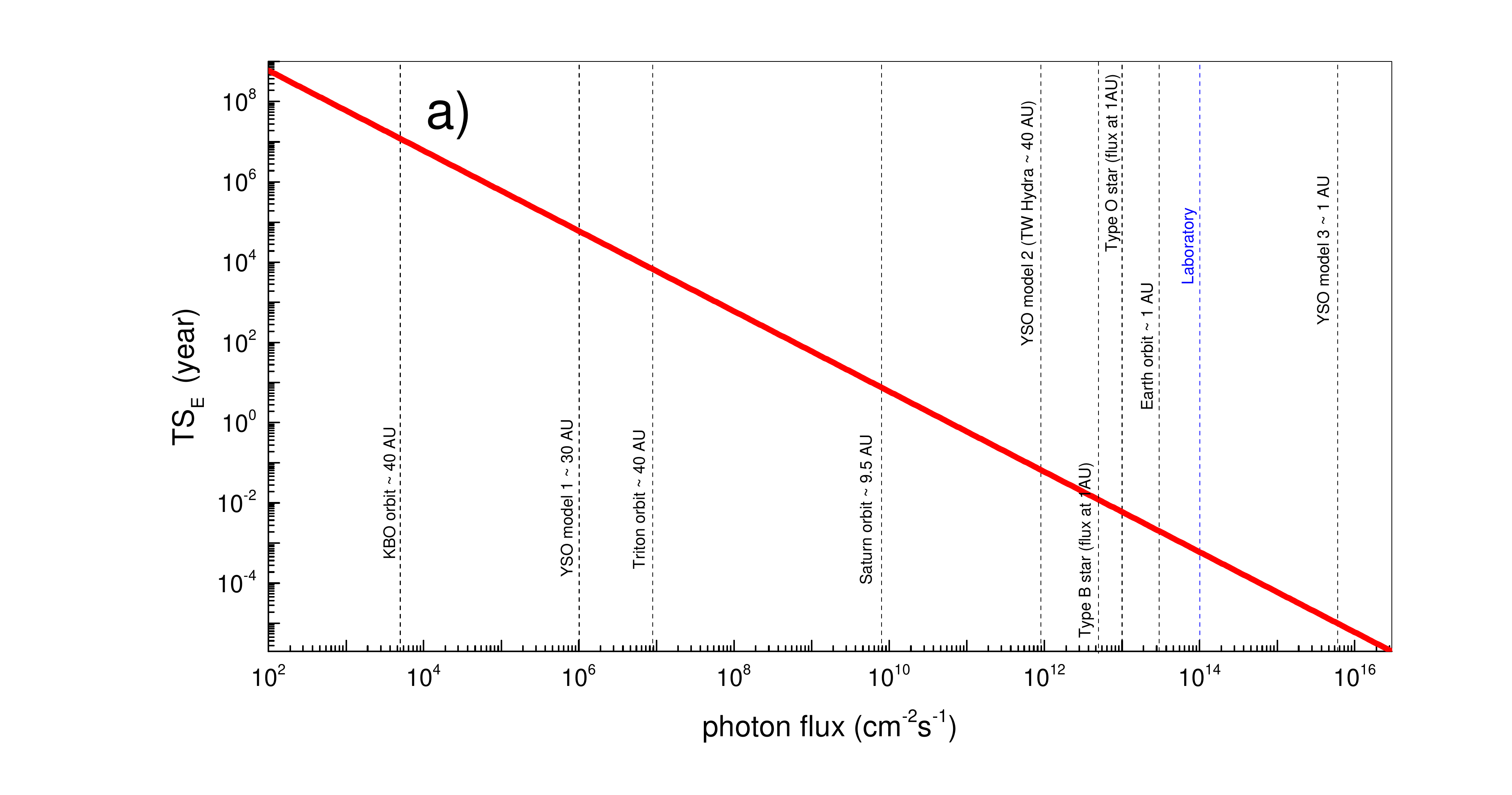}
 \vspace*{-0.9cm}
 \includegraphics[height=7.8cm, width=0.85\textwidth]{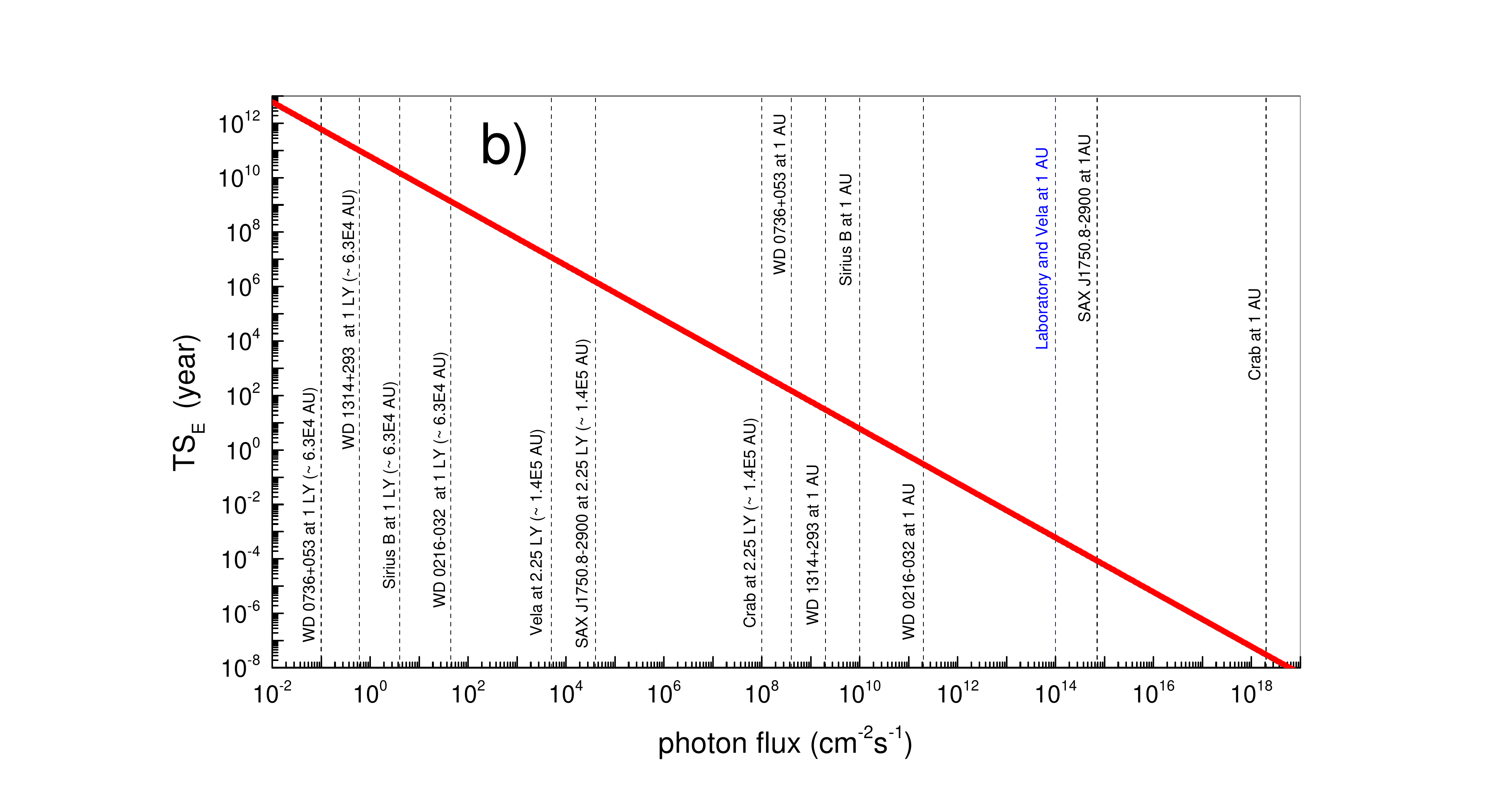}
 \includegraphics[height=7.8cm, width=0.85\textwidth]{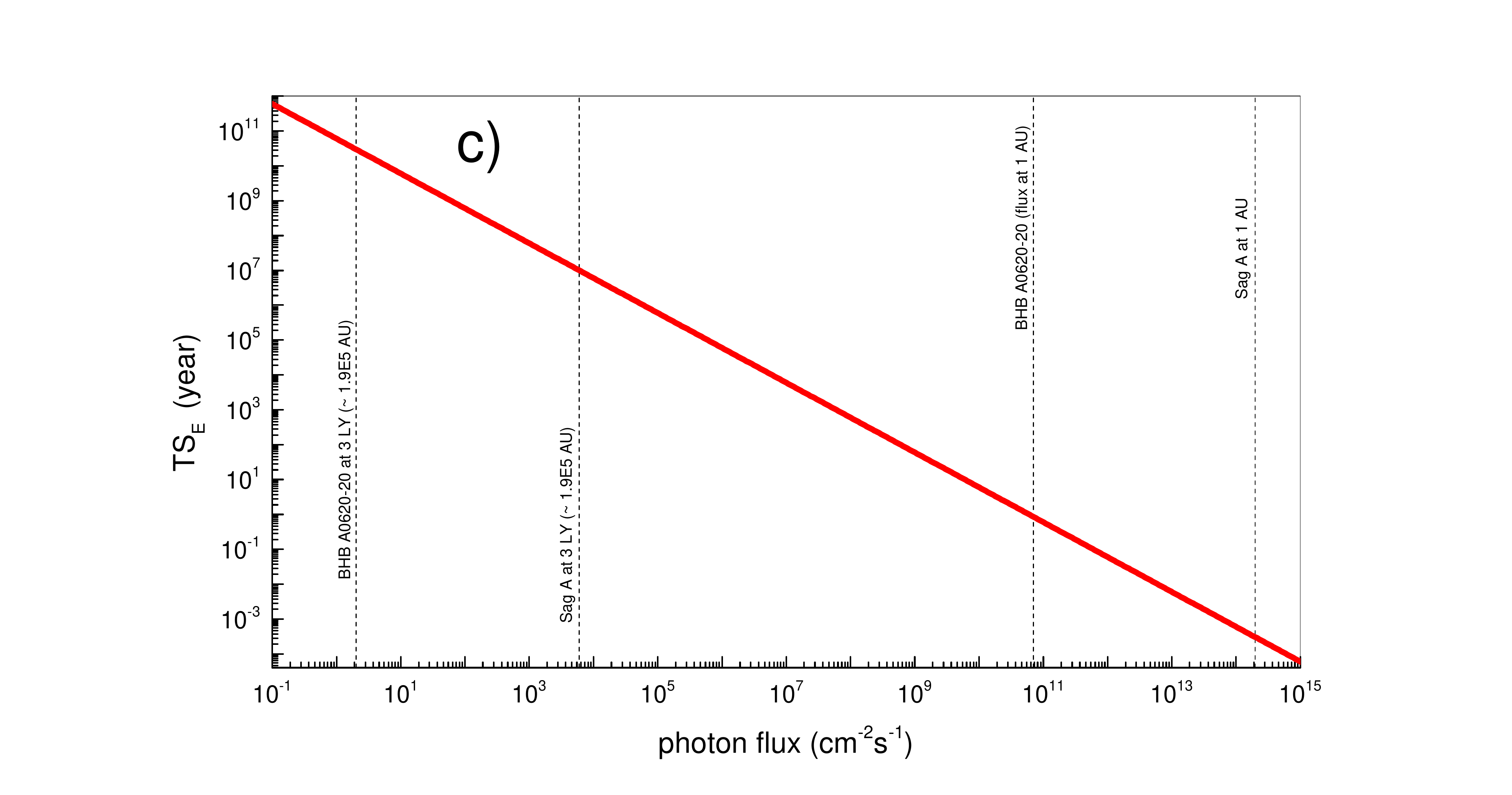}
 \caption{\label{timescale} The timescale to reach chemical equilibrium of typical astrophysical ices exposed to X-rays as function of photon flux (from roughly 6 to 2000 eV). Vertical lines mark the soft X-ray flux of selected space environments at specific distance as listed in Table \ref{photonfluxes}. a) Young stars and main-sequence stars, b) white dwarfs and neutron stars, and c) black holes.}
\end{figure*}

The magnetic field of WDs can be very high and the most magnetic observed white dwarf possesses a magnetic field of $10^9$ G. This high magnetic field implies a huge energy enclosed in a small region of space, which means a potential mechanism of generation of high energy radiation, such as X-rays \citep{Rueda2019Mar}. Curiously, WD stars emit X-rays mainly in binary systems. For binaries, the accretion mechanism is the main responsible for production of radiation; although, there are theoretical models that predict X-ray emission from rotation powered mechanisms. We take a sample of 4 white dwarfs in binary systems: WD 1314+293 that possesses a red dwarf companion, WD 0216-032 with a red giant companion, WD 0736+053 which has a main-sequence star as its companion and Sirius B which is the closest white dwarf from Earth. Table \ref{photonfluxes} shows estimates of photon flux and its corresponding timescale to reach chemical equilibrium. Snow line distances employed here are a rough estimate based on the fact that most planetary nebula has a typical size of a few light-years (LY), so, the 1 LY distance is taken as a fiducial value for WDs snow region where molecules in very low temperatures could be found, thus forming ice grains. Considering this value of distance the timescale of chemical equilibrium is between $1.4\times 10^9 - 6 \times 10^{11}$ years for objects we listed above, see also panel b of Fig. \ref{timescale}. In Table \ref{photonfluxes} we took the standard distance - 1 astronomical unit (AU) - and the snow line distance - 1 light-year (LY) - to calculate photon fluxes and timescale of chemical equilibrium. 

Neutron  stars  are  also  a  strong  X-ray  (thermal  and  non-thermal)  emitter. Here, we selected 3 neutrons stars to characterize their X-ray ionizing field: the young Crab pulsar, the older Vela pulsar and the quiescent emission of the binary SAX J1750.8-2900. The Crab and Vela pulsar are isolated neutron stars which have a soft X-ray spectrum. Considering this, we employed equation \eqref{phifox} to calculate photon fluxes as a function of distance by using X-ray luminosities extracted from \citep{Wijnands2013Jul,Parikh2017Aug,Vanderbosch2019Aug}. The Crab and Vela pulsars are part of  their correspondent nebulae. Crab nebula’s size is around 5 LY. Since the lowest temperature of the nebula would be in its edge, we assume a snow line distance of 2.25LY for the Crab pulsar. We then consider as a typical distance of ice grains in the vicinity of neutron stars the 2.25 light-years one, which corresponds to the radius of the Crab nebula (a supernova remnant). The edge of the Crab nebula could contain molecules in ice phase for what the timescale of chemical equilibrium would be hundreds of years. The obtained values for the timescale for chemical equilibrium of ices in the vicinity of objects like pulsars and the SAX binary  are presented in the panel b of Fig. \ref{timescale} and listed in Table \ref{photonfluxes}. It is worth to cite that many pulsars have been observed to possess planetary objects orbiting it, so, timescales of chemical equilibrium obtained in this work are applicable to those situations, where one needs only  to employ our data for fluence of equilibrium combined with data of distances and photon fluxes of the pulsars. For example, we estimate a timescale of $\sim$600 years for a typical astrophysical ice in the outer border of Crab nebulae to reach chemical equilibrium due to the X-rays emitted by the central neutron star. In the case of the neutron star at Vela nebula at the distance of 2.25 light-years the timescale to reach chemical equilibrium of eventual ices is much higher, around $1.2 \times 10^7$ years.

Neutron stars in binary systems can have non-thermal radiation due to accretion, so the spectra are changed presenting now a hard X-ray component. Since the X-ray spectra range from 10 eV to 10 keV, where the hard X-ray has a power-law shape with peak around 1 keV we take for binary systems one third of the X-ray luminosity value to estimate photon flux. Due to similar reasons, the same procedure is employed to estimate the photon flux generated by binary and supermassive black holes. To convert energy to photon number, and hence luminosity to photon flux, we take the representative value of frequency $\sim$1keV - as performed before, once the majority of X-ray spectrum of the considered objects has peak position around this value – together with equation \eqref{phifox}. Changes on this value of representative frequency or on the partial value of luminosity we take for binaries and black holes do not change drastically the timescale of chemical equilibrium (the timescales may vary only one order of magnitude).

\begin{figure*}
\centering
 \includegraphics[height=13cm, width=17cm]{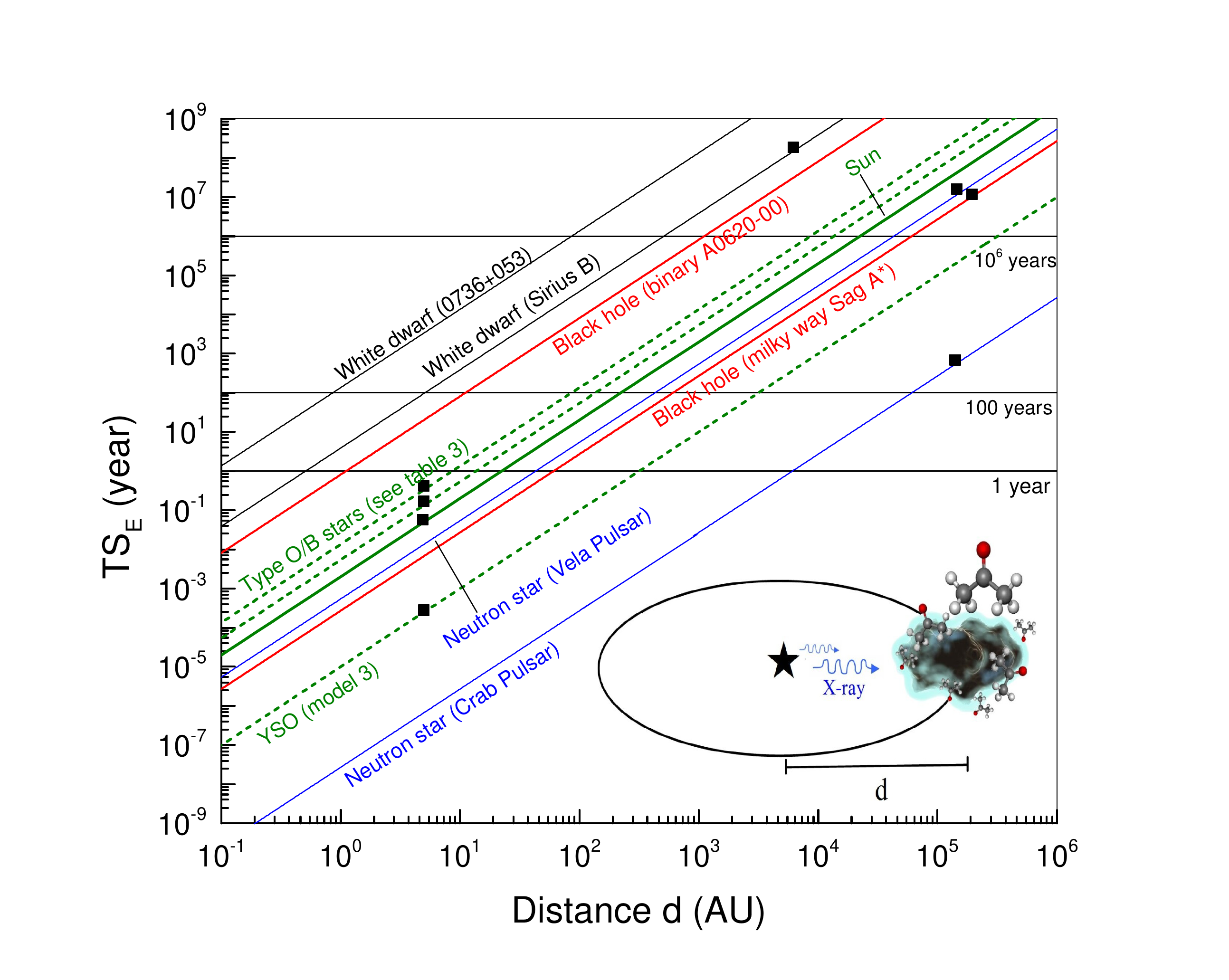}
 \caption{\label{timescaleXdistance} Timescale to reach chemical equilibrium of typical astrophysical ices exposed to X-rays as a function of the distance of the radiation source. The sloped lines with different colors indicate the calculation for different objects (black: white dwarfs, red: black holes; green: OB stars, sun and young stars; blue: neutron stars). The horizontal lines mark the timescales of 1, 100 and 10$^6$ years. Black squares indicate the estimated snow line distance for each object. The inset figure illustrates the irradiation of ice in the vicinity of X-ray source.}
\end{figure*}

\subsection{Black holes}

In this manuscript, we take two representative BHs to characterize their X-ray field:  A0620-00 and Sagittarius A*. The A0620-00 is a binary black hole that presents a quiescent X-ray emission and Sagittarius A* is the supermassive black hole at the center of our galaxy. Recently, a new proposed mechanism could yield the formation of planets around supermassive black holes \citep{Wada2019Nov}. Most of galaxies host a supermassive black hole at their centers, with masses that ranges from millions to billions solar masses. Accretion of gas onto the supermassive black holes is thought to be the main engine of active galactic nuclei. In the midplane, after the accretion disk, a dust torus is formed, where cold, dense gas forms a thin disk. Thus, beyond this snow line ice dust particles can be present. The ice grains aggregates evolve by collisions to form planetesimals. The dust torus extends to around 0.3-30 light-years \citep{Wada2019Nov}. Thus, we consider for black holes a particular distance of 3 light-years as the hypothetical snow line distance. Sagittarius A* has a relatively low quiescent emission compared to other supermassive black holes, and thus its photon flux is only $6\times 10^{3}$ cm$^{-2}$ s$^{-1}$ at 3 light-years distance, and the corresponding timescale for this flux is of the order of $10^7$ years. However, many supermassive black holes have much larger X-ray luminosity of order $10^{12}$ times greater than the Sagittarius A* one, and, hence, could have a smaller timescale for reaching chemical equilibrium. Fig. 6c illustrates the typical timescales of chemical equilibrium for ices in the torus disk in the vicinity of these selected black holes. It is worth to clarify that we take equation \eqref{phifox}, along with one third of X-ray luminosities due to the hard X-ray components of those objects, for both selected BHs to estimate photon fluxes. 

\subsection{Timescale to reach chemical equilibrium}

As we discussed previously, at large fluence (and at constant temperature) the irradiated ices reach chemical equilibrium and its molecular abundances remaining unaltered. This stage happens first at the called Fluence of Equilibrium $F_E$, and with this value taken from the laboratory experiments, we can estimate the timescale of astrophysical ice analogs needs to reach chemical equilibrium in selected astrophysical environments by using the following formulae
\begin{equation}
  TS_{\rm E} = 3\times 10^{-8} \times \frac{F_{\rm E}}{\phi} ~~ {\rm [years]},
\end{equation}
where $F_{\rm E}$ is the equilibrium fluence (for the current ice is around $\sim 2.5\times 10^{18}$ photons per cm$^2$) and $\phi$ is the X-ray flux in the range $6-2000$ eV, in units of photons cm$^{-2}$ s$^{-1}$.

Fig. \ref{timescale} shows the timescale for chemical equilibrium ($TS_{\rm E}$) of typical astrophysical ice (considering here the acetone ice at 12 K) as function of the incoming soft X-ray flux.  Vertical lines indicate the soft X-ray flux of selected space environments. Panel a) shows values for the main-sequence stars and young stars, b) white dwarfs and neutron stars, and c) black holes.

The timescale to reach chemical equilibrium of typical astrophysical ices exposed to X-rays as a function of the distance of the radiation source is given in Fig. \ref{timescaleXdistance}. The sloped lines with different colors indicate the calculation for different objects (black: white dwarfs, red: black holes; green: OB stars, sun and young stars; blue: neutron stars). In estimating photon flux we consider it scales with distance ($\sim d^{-2}$), i.e., the geometric factor is the only responsible for attenuating photon flux. The horizontal lines mark the timescales of 1, 100 and 10$^6$ years. The inset figure illustrates the irradiation of ice in the vicinity of X-ray source. For example, in the case of the Crab pulsar the one year timescale corresponds to a distance in the order of 10$^3$ AU. The distance is an important parameter since it diminishes the photon flux, thus enhancing timescale to reach chemical equilibrium. Black squares indicate the estimated snow line distance for each object (for WD0736+053 and binary BH A0620-00 the snow line distance is out of range of this figure also indicating a very long $TS_{\rm E}$ for ices in this location). From this figure, we can observe that white dwarfs have larger timescales with smaller distances due to also smaller X-ray luminosities. The larger distances are needed for neutron stars, black holes and YSO model 3 because of their high X-ray luminosities.

\section{Conclusions}\label{sectV}

In this work, we consider acetone molecule as a representative complex organic molecule (COM) present on interstellar ice grains under the influence of X-ray radiation. We perform an experimental investigation of the photolysis induced mainly by soft X-rays (from the Brazilian synchrotron light source LNLS/CNPEM) in astrophysical ice analogs (frozen acetone at 12 K). The ice was monitored {\it in-situ} with infrared spectroscopy under ultra-high vacuum conditions. The employed methodology allowed us to simulate the chemistry triggered by photochemical processes in cold, frozen space environments, such as dust torus around supermassive black holes, outer solar system ice bodies, planetary debris close to white dwarfs, planets orbiting pulsars, and so on. This work is the first attempt in the literature to characterize the processing of organic-rich ices astrophysical analog by soft X-rays in the vicinity of compact objects and at specific distances (e.g. snow line region). 
The main conclusions can be summarized as the following:
\begin{enumerate}
    \item The effective destruction cross sections of frozen acetone by broadband soft X-rays, considering different vibration modes, were in the range between $10^{-17} - 10^{-18}$ cm$^{2}$. The most sensitive vibration mode to the incoming radiation was the -C-O band ($\sigma_d = 1.1\times 10^{-17}$ cm$^{2}$).
    \item The photolysis of acetone ice yields the formation of new molecules, such as the C$_3$H$_8$O, CO$_2$, H$_2$CCO and CH$_4$ with an effective formation cross-section in the range from $\sim 10^{-18}$ to $\sim 10^{-19}$ cm$^{2}$. Among the identified species CH$_4$ followed by H$_2$CCO were the most abundant.
    \item The chemical equilibrium within the ice sample were obtained at the fluence around $F_E = 2 \times 10^{18}$ photons cm$^{-2}$. Molecular abundances at this stage were estimated (e.g. unknown species $\sim 14.3\%$; acetone $\sim 79.1\%$; H$_2$CCO $\sim 1.5\%$; CH$_4$ $\sim 4.3\%$; CO$_2$ $\sim 0.8\%$).
    \item Timescale to reach chemical equilibrium in selected space environments (e.g. near some young stars, main-sequence stars, and compact objects including WD stars, neutron stars and BHs) as a function of photon flux and as a function of distance from the central radiation source was calculated. For compact objects, we have found that depending on the emission, the timescale can be either very small or for a reasonable timescale the distance needs to be very large (1-3 light-years), the second scenario is more physically feasible since ice grains and planetesimals would form only at large distances from the central X-ray source. The timescales for ices to reach chemical equilibrium in hypothetical snow line distances from the selected studied X-ray sources were given, e.g. we estimate timescales of 18 days, 3.6 and 1.8 months, $1.4\times 10^{9}-6\times 10^{11}$ years, 600 and $1.2\times 10^7$ years, and $10^7$ years, for the Sun at 5 AU, O/B stars at 5 AU, for white dwarfs at 1 LY, for the Crab pulsar at 2.25 LY, for Vela pulsar at 2.25 LY, and for Sagittarius A* at 3 LY, respectively.  
\end{enumerate}

The current work shows good similarity with other previous experiments employing X-rays on water-rich ices containing organic species \citep[e.g.][]{Pilling2015Sep,deA.Vasconcelos2017Nov,Bonfim2017Mar,Pilling2019}. This study also helps our current understanding about how ionizing radiation affects the chemistry of frozen material in space, for example, in outer solar system bodies and in the vicinities of young stellar and compact objects.

\section*{Acknowledgements}
The authors acknowledge the FVE/UNIVAP and the Brazilian research agencies CNPq (projects \#304130/2012-5; \#306145/ 2015-4; \#302985/2018-2), FAPESP (projects JP 2009/18304-0, DR 2013/07657-5, PD 2015/10492-3). The authors also acknowledge the staff of LNLS/CNPEM and UNIVAP for their invaluable support and thank the Coordena\c{c}\~ao de Aperfei\c coamento de Pessoal de N\'ivel Superior (CAPES) grant PNPD/88887.368365/2019-00. Authors thank Mrs. Pel\'ogia A. B. R. for the english revision.

\section*{Data availability}

The data underlying this article will be shared on reasonable request to the corresponding author.




\bibliographystyle{mnras}
\bibliography{references} 








\bsp	
\label{lastpage}
\end{document}